\documentclass{article}

\usepackage{amsfonts}
\usepackage{xspace} 
\usepackage{paralist}
\usepackage{url}

\usepackage{tikz}
\usetikzlibrary{backgrounds}

\newcommand{\G}{\ensuremath{\mathbb{G}}\xspace}
\newcommand{\Zq}{\ensuremath{\mathbb{Z}_q}\xspace}

\title{Verifiable Elections with \\ Commitment Consistent Encryption\\ {\Large A Primer}}
\author{Olivier Pereira}
\date{August 15, 2014}

\begin{document}
\maketitle

\begin{abstract}
  This note provides an introduction to the PPATS Commitment
  Consistent Encryption (CCE) scheme proposed by Cuvelier, Pereira and
  Peters~\cite{CPP13} and its use in the design of end-to-end
  verifiable elections with a perfectly private audit trail. These
  elections can be verified using audit data that will never leak any
  information about the vote, even if all the private keys of the
  elections are compromised, or if the cryptographic assumptions are broken.
\end{abstract}

\section{Introduction}
\label{sec:motivation}

The verification process of end-to-end verifiable elections includes
the examination of a set of audit data made available, usually through
a public website, to anyone who would like to verify the
election. Once these audit data are released, they can be copied by
numerous people around the world and stored for ever. In order to make
sure that these data \emph{hide} the votes but still \emph{bind} the votes to the
election result, that is, prove that this result is correct, they are
typically produced using various cryptographic techniques.

In most end-to-end verifiable voting schemes, these hiding and binding
properties of the audit data are not absolute. They rely on the
assumption that pieces of secret information held by some trustees
remain secret and on the assumption that solving some mathematical
problems cannot be done because it would require way too much
computational power. This is the case, for instance, when the audit
data contain votes that are encrypted with a public key encryption
scheme (ElGamal, Paillier, \dots), or are hidden thanks to the use of
a pseudorandom generator, or when the Fiat-Shamir heuristic is used to
produce efficient non-interactive validity proofs. 

The two flavors in which non absolute security properties appear, that
is, depending on keeping secrets and depending on computational
assumptions, are actually fairly different. Stealing secret
information is something that may happen at any time or, hopefully,
will not happen at all. But, if it happens, it is likely to happen
during the election, as this is when the secrets are manipulated ---
they should normally be destroyed after that.  By contrast,
computational advances are expected to happen no matter what, but we
can try to estimate their pace and reasonably hope that they will not
happen before several years.

In some cases though, we can design schemes that have absolute or
perfect security properties. We know that these schemes will never be
broken, independently of any stolen keys or computing
advances. Unfortunately, we also know that, under natural settings, we
cannot have a voting scheme that offers the perfect flavors of the
binding and hiding properties.

Let us now consider the consequences of stolen secrets and
computational advances on our hiding and binding properties. Regarding
the binding property, all end-to-end verifiable systems are designed
in such a way that the correctness of the result does not depend on
secret information kept by trustees: these trustees would de facto be
able to fake the election results, which is far more trust than we are
ready to give them. The impact of computational progresses on the
binding property still exists, but is much more benign: if some
advances make it possible, in several years, to create new audit data
that would pass an election verification procedure and support a
different result, such data will not convince anyone since we will
know that the strength of the cryptographic schemes on which the
verification procedure relies is gone. Furthermore, the election
results would have been validated for a long time, and it would be
impossible to change anything anyway. So, proofs must be available when
they are needed, but may become unconvincing in the future, when they
do not matter anymore.

Regarding the hiding property, stolen secret information would
typically result in the loss of privacy of the votes, which certainly
is very damaging.
The eventual compromise of the privacy of the votes due to
computational advances is also a problematic threat: this perspective
can be sufficient to enable coercion, which is the main threat that
the secret ballot is expected to prevent.

The discussion above is summarized in Table~\ref{tab:
  properties}. This table suggests that targeting a perfect flavor of
the hiding property is quite appealing, as it would eliminate the
threats from the last line of the table. This imposes keeping a
computational flavor of verifiability, but this looks acceptable as
long as this verifiability does not rely on keeping secrets. These are
the properties we are targeting in this note, which are also summarized
in~Table~\ref{tab: properties-PPAT}.

\begin{table}[ht]
  \centering
  \begin{tabular}[c]{|r|c|c|}
\hline
                    & compromise of secret & computational advances \\
                    & information & in the future \\
\hline
  Binding breaks due to:   & Cannot happen in  & Benign effect \\
                           & E2E verifiable systems & \\
\hline
  Hiding breaks due to:    & Highly             & Damaging  \\
                    & Damaging         & and unavoidable \\
\hline
  \end{tabular}

\caption{Effect of cryptographic issues on the binding and hiding properties of public audit data in most end-to-end verifiable election systems.}
\label{tab: properties}
\end{table}

\begin{table}[ht]
  \centering
  \begin{tabular}[c]{|r|c|c|}
\hline
                    & compromise of secret & computational advances \\
                    & information & in the future \\
\hline
  Binding breaks due to:   & Cannot happen  & Benign  effect\\
\hline
  Hiding breaks due to:    & Cannot happen             & Cannot happen  \\
\hline
  \end{tabular}

\caption{Effect of cryptographic issues on the binding and hiding properties of public audit data when using a perfectly private audit trail.}
\label{tab: properties-PPAT}
\end{table}

\section{Commitment-Consistent Encryption}
\label{sec:hiding-votes-binding}

Commitment-consistent encryption (CCE) schemes~\cite{CPP13} are
cryptographic mechanisms that can provide verifiable elections with
the properties described in Table~\ref{tab: properties-PPAT}.

In many large scale verifiable election schemes, votes are transmitted
to a set of trustees under the protection of a public key encryption
mechanism.  However, the ciphertexts produced by a public key
encryption mechanisms cannot be perfectly hiding and, therefore,
cannot be part of our election audit data.

Commitment-consistent encryption schemes are public key encryption
sche\-mes with an extra feature that addresses this issue: from any
encrypted vote, it is possible to extract a perfectly hiding
commitment that is consistent with that vote. That commitment can be
part of the audit data, be verified by anyone, and we are sure that it
cannot be manipulated without breaking the binding property of the
commitment scheme.

\subsection{Computational setting}
We are looking for efficient instances of CCE schemes that can be
easily used in a distributed setting. For this reason, the schemes
that we discuss here work with elements of public prime order
groups. This in particular enables efficient distributed key
generation~\cite{GJKR07} and the manipulation of integers that are shorter than
those occurring in schemes that rely on the hardness of factoring.

So, we are going to compute in cyclic groups of prime order
$q$. We refer to these groups with the letter \G, possibly indexed, and
use the letters $g$ and $h$  to denote generators of such
groups.

The computational security of our schemes is guaranteed as long
as the Decisional Diffie-Hellman (DDH) problem~\cite{B98} is
hard. Informally, this problem consists in deciding whether a tuple is
sampled from the distribution $(g, g^x, g^y, g^{xy})$ or from the
distribution $(g, g^x, g^y, g^z)$ for uniformly random $(x, y, z)
\leftarrow \Zq^3$. It is believed to be hard in large prime order
subgroups of $\mathbb{Z}_p^*$ and in various groups obtained from
elliptic curves for instance. The hardness of this problem implies the
hardness of other related problems, including the discrete logarithm
(DL) problem, that states that is hard to compute $x$ from inputs $(g,
g^x)$ where $x$ is uniformly random in \Zq.

\subsection{Perfectly hiding commitments}
\label{sec:perf-hiding-comm}

One of the most popular example of perfectly hiding commitment scheme
was introduced by Chaum et al.~\cite{CDG87}, and is often called the
Pedersen commitment: given two random generators $g_1$ and $g_2$ of
the group $\G_1$,\footnote{Such generators can be obtained through a
  mapping from the outputs of a PRG or from the digits of a public
  constant like the number $\pi$.} we can commit on a vote $v \in
\Zq$, by computing the value $c=g_1^rg_2^v$ for a random element $r
\leftarrow \Zq$. An opening of this commitment is simply made of the
values $r$ and $v$ (but we later omit this last value, and assume that
it can be efficiently computed by exhaustive search, which is
certainly the case when $v$ is a single bit). This scheme is
perfectly hiding: since $g_1^r$ is a uniformly random group element,
the resulting $c$ is also uniformly distributed in $\G_1$ (this is
equivalent to running a one-time pad in $\G_1$). It is also binding if the
DL problem is hard in $\G_1$: any two distinct openings $(r_1, v_1)$
and $(r_2, v_2)$ of the same $c$ would immediately lead to computing
the discrete logarithm of $g_2$ in basis $g_1$, which would be equal
to $\frac{r_2 - r_1}{v_1 - v_2}$.

These commitments are quite appealing: they have a simple expression
and are homomorphic: the product of two commitments can be opened with
the sum of the committed values and randomnesses. This property is
most useful, as it makes it possible to multiply committed votes in order
to obtain a commitment on the sum of these votes.

In order to make a CCE scheme based on this commitment scheme, we
would then need to encrypt $r$ (and possibly $v$ as well) with an
additively homomorphic encryption scheme, so that the trustees would
be able to compute an opening of the product of the commitments on all
votes. Unfortunately, traditional additively homomorphic encryption
schemes have either exponential decryption time, which is not
acceptable given the large size of $r$, or require using groups of
composite order, which we want to avoid for the reasons described
above.

Recently, another way of opening these commitments was proposed by Abe
et al.~\cite{AHO12}, and can serve our purpose. The proposal is to use
a second group $\G_2$ of same prime order $q$, with generator $h_1$,
and to define the opening of a commitment $c=g_1^rg_2^v$ as $a=h_1^r$:
computing that last value indeed seems to require the knowledge of
$r$. However, we need a mechanism to be able to verify that $a$ is
really an opening of $c$ for a given vote $v$. To this purpose, $\G_1$
and $\G_2$ are chosen to be groups admitting an efficient bilinear map
$e: \G_1 \times \G_2 \rightarrow \G_T$, where $\G_T$ is called the
target group.\footnote{Choices for the groups $\G_1$ and $\G_2$ are
  based on elliptic curves admitting a so-called Type~3 (asymmetric)
  pairing~\cite{GPS08}, which provide the most freedom and the most
  efficient implementations in the pairing landscape. It is not
  difficult to transpose the schemes we propose to the symmetric
  setting (Type~1 pairing), but the resulting solutions are less
  efficient and cannot be based on the DDH problem anymore, since DDH
  is easy to solve in these groups (the DLIN problem is a natural
  choice there.)} The bilinearity here implies that, if $e(g_1, h_1) =
g_T$, then $e(g_1^a, h_1^b) = g_T^{ab}$. Thanks to this bilinear map,
it is possible to verify that $a$ opens the commitment $c$ for vote
$v$ by checking that $e(c/g_2^v, h_1) = e(g_1, a)$.

These commitments certainly remain perfectly hiding, for the same
reason as above. The binding property holds if the DDH problem is hard
in $\G_1$. Suppose indeed that we get a DDH challenge tuple $(g_1,
g_1^x, g_1^y, g_1^z)$ and need to decide whether $z = xy$ or not. We
can set $g_2 = g_1^x$ and feed an adversary against the commitment
scheme with the bases $(g_1, g_2, h_1)$. This adversary will provide
us with two distinct openings $(v_1, a_1)$ and $(v_2, a_2)$ of a
commitment of its choice. We can now verify that $z = xy$ if and only
if $e(g_1^z, h_1) = e(g_1^y, (a_1/a_2)^{\frac{1}{v_2 - v_1}})$.

So, we have a scheme that provides commitments that are as efficient
to compute as Pedersen commitments, but with openings that are group
elements. The opening verification operation becomes more expensive,
though, as the pairing operation is considerably more demanding than
an exponentiation in one of the base groups. However, this extra cost
remains low at an election scale as, when using homomorphic tallying
techniques, we will only need to verify one commitment opening per
question, and not per voter. Furthermore, since all computational
operations are performed with public bases and secrets in the
exponents, this scheme is fully compatible with the most efficient
sigma proof techniques.

\subsection{The PPATS scheme}

In order to have an election with a perfectly private audit trail, we
ask voters to commit on their votes using the scheme we just
described. Talliers are then expected to publish the election
result, and show that they can open the product of the committed votes
to this result: this is a simple way to prove the correctness of the
outcome based on the binding property of the commitment scheme.

The talliers then need a way to compute that opening, which simply is
the product of the openings of the individual votes. To this purpose, the voters
also  send the opening of their individual votes to the
talliers, protected using an additively homomorphic threshold
encryption scheme in order to make sure that no subset of the talliers
smaller than the chosen threshold would be able to obtain information
about individual votes. ElGamal encryption can be used for that
purpose and pairs nicely with the commitment scheme, as its security
also relies on the hardness of the DDH problem.

Protecting commitment openings with encryption can place the trustees
in a difficult position, though: they become unable to determine
whether the encrypted information that is sent to them really contains
an opening of a vote commitment, or just an arbitrary value. The
second case could lead to a very uncomfortable situation, as the
trustees would become unable to compute the election results, and
would have a hard time proving that they are acting in good faith
without violating or reducing the privacy of honest voters. A simple
solution to this issue is to require voters to prove the consistency
of the CCE ciphertexts they produce. This can be done using a sigma
proof, thanks to the structure of our commitment openings.

The resulting scheme is called the PPATS encryption scheme, and is
described in Table~\ref{tab:PPATS} (for efficiency reasons and
consistency with the literature, we switch the roles of $\G_1$ and
$\G_2$ compared to our text above: operations in $\G_2$ are usually
more expensive than those in $\G_1$.)

\begin{table}[p]\centering
  
\fbox{
  \begin{minipage}[c]{.9\linewidth}
   \textbf{PPATS encryption} 
   \begin{compactitem}
   \item \emph{Setup:} Select type-3 pairing-friendly groups $\G_1,
     \G_2, \G_T$ of prime order $q$, together with random generators
     $g_1$ of $\G_1$ and $h_1, h_2$ of $\G_2$.
   \item \emph{Key Generation:} Generate an ElGamal public encryption
     key $g_2 = g_1^x$. The secret key is $x$, possibly existing
     under a distributed (threshold) form only.
   \item \emph{Encryption:} Encrypt vote $v$ as $(c_1, c_2, d,
     \sigma_{cc}) = (g_1^s, g_1^rg_2^s, h_1^rh_2^v, \sigma_{cc})$,
     using uniformly random $(r, s) \leftarrow \Zq^2$, and the
     consistency proof $\sigma_{cc}$ described in
     Table~\ref{tab:validity}.
   \item \emph{Decryption:} Extract the discrete logarithm of
     $e(c_1^{x}/c_2, h_1)e(g_1, d)$ in basis $e(g_1, h_2)$. (The
     computation of $c_1^x$ can be done in a distributed
     manner.)
   \item \emph{Extraction of commitment:} The perfectly hiding commitment
     extracted from a ciphertext $(c_1, c_2, d, \sigma_{cc})$ is $d$. 
   \item \emph{Extraction of commitment opening:} The commitment
     opening is computed as $a = c_2/c_1^x$.
   \item \emph{Opening verification:} Given a commitment $d$ and an
     opening $a$ for vote $v$, verify if $e(a, h_1) = e(g_1, d/h_2^v)$. 
   \end{compactitem}
 \end{minipage}
}
\caption{The PPATS encryption scheme}
\label{tab:PPATS}
\end{table}

\begin{table}[p]
  \centering

\fbox{
  \begin{minipage}[c]{.9\linewidth}
  \textbf{PPATS Consistency Proof} for triple $c = (g_1^s, g_1^rg_2^s, h_1^rh_2^v)$
  \begin{compactitem}
  \item \emph{Commitment computation:} Compute $c' = (c'_1, c'_2, d') =
    (g_1^{s'}, g_1^{r'}g_2^{s'}, h_1^{r'}h_2^{v'})$ using uniformly
    random $(r', s', v') \leftarrow \Zq^3$
  \item \emph{Challenge computation:} Compute $e = \mathcal{H}(c, c',
    label)$. Here, $\mathcal{H}$ is a cryptographic hash function and
    $label$ is a global public value that contains the description of
    the groups that are used, together with the generators of $\G_1$
    and $\G_2$ used in the proof. It should also contain some
    identifiers for the election and the purpose of the proof.
  \item \emph{Response computation:} Compute $f_r = r' + er$, $f_s =
    s' + es$ and $f_v = v' + ev$. 
  \end{compactitem}
   The proof is defined as $\sigma_{cc} = (e, f_r, f_s, f_v)$.
 
   \medskip
   \textbf{PPATS Validity Verification} for triple $c = (c_1, c_2, d)$ and proof $\sigma_{cc} = (e, f_r, f_s, f_v)$.
   \begin{compactitem}
   \item \emph{Commitment reconstruction:} Compute $c' = (c'_1, c'_2,
     d')$ as follows: $c'_1 = g_1^{f_s}/c_1^e$, $c'_2 =
     g_1^{f_r}g_2^{f_s}/c_2^e$, $c'_3 = h_1^{f_r}h_2^{f_v}/d^e$.
   \item \emph{Challenge verification:} Verify if $e =
     \mathcal{H}(c, c', label)$.
   \end{compactitem}
   The validity verification returns the result of the test above. 

\end{minipage}
}
  \caption{PPATS Validity Proof}
\label{tab:validity}
\end{table}

The PPATS encryption scheme is additively homomorphic, but its
decryption procedure can be quite slow if the values to be decrypted
are (very) large, since the extraction of a discrete logarithm is
required.  This should not be a problem for elections with homomorphic
tallying (values up to $2^{50}$ can still be extracted in seconds
using the baby-step giant-step algorithm), but may become an issue if
unpredictable write-ins need to be taken into account for
instance. For this purpose, the PPATC encryption scheme~\cite{CPP13}
can be used, which enables efficient decryption and is compatible with
efficient mixnets but is not additively homomorphic anymore.

\section{Voting based on PPATS encryption}
\label{sec:voting-based-ppats}

The PPATS scheme can be conveniently used to build a voting scheme
with homomorphic tallying. We outline the main steps of a simple
voting scheme here. This process can of course be refined with various
standard enhancements (cast or audit procedure, \dots) or adapted to
completely different E2E voting schemes.

\begin{enumerate}
\item Groups are chosen, depending on the security parameter, and made public.
\item Trustees generate a PPATS key in a threshold manner. Any
  protocol (see, e.g., Gennaro et al.~\cite{GJKR07}) that can be used
  for DDH-based cryptosystems can be used for PPATS encryption.
\item A public bulletin board is created, with the election
  description and the PPATS public key (including information that may
  be necessary to verify the validity of this key.)
\item Voters prepare their ballot by producing a PPATS encryption of a
  0 or a 1 for each response to the questions. They also publicly
  prove the validity of their vote by submitting, for the commitment
  included in each ciphertext, a proof of knowledge $\sigma_{0/1}$ of
  an opening of $d$ to a 0 or a 1.\footnote{For commitment $d$, this can be a
  disjunctive proof of knowledge of the discrete logarithm of either
  $d$ or $d/g_2$ in basis 
  $g_1$~\cite{CDS94}. 
}
\item When receiving a ballot, the trustees (or the bulletin board, by
  delegation) check the validity of all PPATS validity proofs and, if
  they check, the bulletin board publishes the commitments extracted
  from these ciphertexts together with the 0/1 validity proofs on these
  commitments. 

\item At the end of the election day, the trustees:
  \begin{enumerate}\item 
    check the validity of all proofs posted on the bulletin board. If
    the proofs check, they
  \item multiply the PPATS ciphertexts from all voters, response by
    response (removing the consistency proofs), obtaining one
    ciphertext per response encrypting the election results. Then, they
  \item extract and publish the commitment opening for each of the
    resulting ciphertext, and publish the decryption of the election
    results. Eventually, they 
  \item erase all secret keys. 
  \end{enumerate}
\item Voters can verify the election as follows: 
  \begin{enumerate}
  \item Check the public parameters of the election (cryptographic
    parameters, questions, voter list, \dots);
  \item Check the proper generation of the election public key; 
  \item Check the validity of all votes committed on the bulletin board;
  \item Multiply the commitments from all voters, response by
    response, obtaining one commitment per response, committing on the
    election results;
  \item Check that the openings provided by the trustees match these
    commitments.
  \end{enumerate}
\end{enumerate}

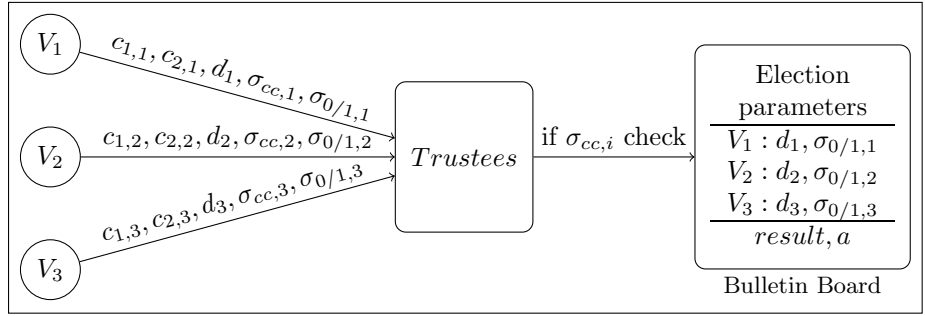
\begin{figure}[ht]
  \centering
  \begin{tikzpicture}[framed]
    \node[draw, circle] (v1) at (0,1.5) {$V_1$};
    \node[draw, circle] (v2) at (0,0) {$V_2$};
    \node[draw, circle] (v3) at (0,-1.5) {$V_3$};

    \node[draw, minimum height=2cm, inner sep=6pt, rounded corners] (t) at (5.5,0) {$Trustees$};

    \draw[->] (v1) -- node[above=-2pt, sloped] %
              {$c_{1,1}, c_{2, 1}, d_1, \sigma_{cc, 1}, \sigma_{0/1, 1}$} (t);
    \draw[->] (v2) -- node[above=-2pt, sloped] %
              {$c_{1,2}, c_{2, 2}, d_2, \sigma_{cc, 2}, \sigma_{0/1, 2}$} (t);
    \draw[->] (v3) -- node[above=-2pt, sloped] %
              {$c_{1,3}, c_{2, 3}, d_3, \sigma_{cc, 3}, \sigma_{0/1, 3}$} (t);

    \node[draw, minimum height=2cm, inner sep=6pt, rounded corners] (bb) at (10,0) %
        {
          \begin{tabular}{c}
            Election\\
            parameters\\ \hline
            $V_1: d_1, \sigma_{0/1, 1}$\\
            $V_2: d_2, \sigma_{0/1, 2}$\\
            $V_3: d_3, \sigma_{0/1, 3}$\\ \hline
            $result, a$
          \end{tabular}
        };
    \node at (10, -1.7) {\small{Bulletin Board}};

    \draw[->] (t) -- node[above=-2pt] {if $\sigma_{cc,i}$ check} (bb);
     
  \end{tikzpicture}
  \caption{Voting process representation for a single question, based
    on PPATS. $V_1$, $V_2$ and $V_3$ are voters. On the bulletin
    board, $result$ is the decryption of $c = (\prod c_{1,i}, \prod
    c_{2,i}, \prod d_i)$ and $a = \prod c_{2, i} / \prod c_{1,
      i}^x$. }
  \label{fig:election}
\end{figure}

This process is outlined and described in terms of the PPATS scheme in Figure~\ref{fig:election}.

This voting scheme satisfies the properties we were looking for. We
have E2E verifiability in the traditional sense: the correctness of
the outcome does not depend on any trustee, or on any secret
information held by anyone. It does depend on computational
assumptions, though: fake results could pass the audit procedure if
the hash function used in the ZK proofs does not properly emulate a
random oracle, or if the DDH problem happens to be easy. These
assumptions are well-known and used in numerous other cryptographic
schemes. Regarding the confidentiality of the votes, we see that the
content of the bulletin board is perfectly hiding: it only contains an
encryption key that is independent of the votes, perfectly hiding
commitments and perfect zero-knowledge proofs for all votes and an
opening of the public election results. So, these audit data cannot
help an adversary that would gain possession of the keys held by the
trustees, or would be able to solve computational problems that are
believed to be hard today.

\section{Security Parameters and Efficiency Notes}
\label{sec:impl-notes}

\paragraph{Group selection.} The most standard choice of curves
admitting a type-3 pairing at the 128 bit security level is the BN
curves~\cite{BN05}, with 256 bit group order. Various implementations of these
curves are available, including in the PBC library~\cite{PBC}, the
MIRACL cryptographic SDK~\cite{MIRACL}, and the RELIC
toolkit~\cite{Relic}.

To provide a rough idea of running times, the benchmarks of the MIRACL
SDK indicate for these curves that exponentiations take .22 ms in
$\G_1$, .44 ms in $\G_2$, and that the pairing operation takes 20 ms
(all on a 2010 Intel i5 520M processor). If we focus on the cost of
exponentiations when computing a ballot (which is the dominant
factor), we can count that a PPATS encryption of a choice together
with a 0/1 proof for the commitment takes 6 exponentiations in $\G_1$
and 5 exponentiations in $\G_2$, for a total computing time of 3.52
ms. So, a modern processor should be able to encrypt around 280
responses per second using a single thread, without any
precomputation. 

\paragraph{Precomputation.} Most of the computational work can be
performed out of critical moments, which can be useful in 
elections with large ballots. For instance:
\begin{itemize}
\item All the exponentiations needed for computing a PPATS ciphertext
  are independent of the actual choices made by the voter. A voting
  client can then perform these operations in advance, while the voter
  makes his selections for instance.
\item All the exponentiations that are needed for ballot preparation
  and proof verification are in fixed public bases, enabling the
  efficient use of various precomputation methods, that usually
  provide a speedup factor between 2 and 3 for usual parameters.
\item The validity of all proofs can be checked during the election
  day: there is no need to wait until the closing of the polls. 
\item The PPATS ciphertexts can be multiplied together as they
  come. In this way, the encrypted results can be available instantly
  when the polls close.
\item The discrete logarithm extraction that is part of the decryption
  of the election results can be immediate if all the powers of the DL
  basis have been precomputed and stored, which is easy for any
  realistic election size.
\end{itemize}

\subsection*{Acknowledgement}
We thank Ron Rivest for encouraging us to write this note.

\bibliographystyle{plain}
\bibliography{PPATBib}

\end{document}